\documentstyle[12pt,epsfig,color]{article}
\begin{document}

\begin{center}
{\Large\bf Spherical collapse in DGP braneworld cosmology}
\\[15mm]
Ankan Mukherjee\footnote{Email: ankan.ju@gmail.com}
\vskip 0.5 cm
{\em $^{1}$Centre for Theoretical Physics,\\ Jamia Millia Islamia, Jamia Nagar, New Delhi-110025, India}\\[15mm]
\end{center}

\pagestyle{myheadings}
\newcommand{\be}{\begin{equation}}
\newcommand{\ee}{\end{equation}}
\newcommand{\bea}{\begin{eqnarray}}
\newcommand{\eea}{\end{eqnarray}}

\begin{abstract}
The DGP (Dvali, Gabadaze and Porrati) braneworld cosmology gives an alternative to dark energy. In DGP cosmology the alleged cosmic acceleration is generated by the modification of gravity theory.  Nonlinear evolution of matter density contrast in DGP braneworld cosmology is studied in the present work. The semi-analytic approach of spherical collapse of matter overdensity is adopted in the present context to study the nonliner evolution. Further, the number count of galaxy clusters along the redshift is studied for the DGP cosmology using Press-Schechter and Sheth-Tormen mass function formalisms. It is observed that for same values of cosmological parameters, the DGP model enhances the number of galaxy cluster compared to the standard $\Lambda$CDM scenario. 
\end{abstract}


\vskip 2.0cm

\section{Introduction}
The observed phenomenon of cosmic acceleration \cite{Riess:1998cb,Perlmutter:1998np,Schmidt:1998ys} and its further confirmations from latest cosmological observations \cite{Suzuki:2012,Crocce:2015xpb,Agathe:2019vsu,Aghanim:2018eyx} have changed our understanding about the universe. 
Within the regime of general relativity (GR), the alleged accelerated expansion of the universe can be explained by introducing an exotic component in the energy budget, dubbed as {\it dark energy}. Besides dark energy cosmology, the other way to look for a possible solution of the puzzle is to modify the theory of gravity where the geometry itself be responsible of the cosmic acceleration. Higher dimensional gravity theories \cite{Dvali:2000hr,Brax:2004xh}, scalar-tensor theories \cite{Bertolami:1999dp,Banerjee:2000mj,Banerjee:2000gt,Sen:2000zk}, Gauss-Bonnet gravity \cite{Nojiri:2005vv,Comelli:2005tn,Andrew:2007xa}, $f(R)$ gravity \cite{Capozziello:2003tk,Carroll:2003wy,Carroll:2004de,Nojiri:2003ft} are amongs the popular modified gravity theories.  Some of the $f(R)$ gravity models are highly successful to explain the cosmic inflation \cite{Starobinsky:1980,Hu:2007nk}. On the other hand, some of the higher dimensional theories \cite{Pahwa:2011td,Alam:2016wpf,Sadeghi:2014jua,Ranjit:2011,Liang:2010jy,Basilakos:2013ij,Biswas:2018mxe,Ravanpak:2019zdg}, are consistent in case of late time cosmology.

The present work is devoted to the study of the evolution of matter overdensity and clustering of dark matter in the higher dimensional DPG (Dvali, Gabadaze and Porrati) braneworld cosmology \cite{Dvali:2000hr}.  The prime endeavour is to probe the effect of higher dimensional gravity scenario on the linear and nonlinear evolution of matter density contrast and formation of large scale structure in the universe. The semi-analytic approach of spherical collapse of matter over density is adopted for the present study. Further the number count of dark matter halos along redshift for the DGP cosmology is also studied. The results are compared with that of standard {\it cosmological constant} with cold dark matter ($\Lambda$CDM) model. It is indispensable for geometric dark energy model to produce the appropriate dynamics of the growth of cosmological perturbations and the large scale structure formation to be accepted as a viable cosmological model. In this regard the present work is important to study the viability of DGP cosmology.  A study of structure formation in DGP cosmology was carried out by Koyama and Maartens \cite{Koyama:2005kd}. Multamaki, Gaztanaga and Manera studied the growth of large scale structure in braneworld cosmology \cite{Multamaki:2003}. Topology of large scale structure in non-standard cosmology was investigated by Watts et al. \cite{Watts:2017lzm}. Large scale structure formation on a normal branch of DGP brane was studied by Song \cite{Song:2007wd}. Numerical study of cosmological perturbation in DGP cosmology was carried out by Cardoso et al \cite{Cardoso:2007xc}. Structure formation in modified gravity scenario is studied by Brax and Valageas \cite{Brax:2012} and by Carrol et al. \cite{Carroll:2006jn}. Aspects of spherical collapse in modified gravity theories are explored by Schafer and Koyama \cite{Schafer:2008} and by Schmidt, Hu and Lima \cite{Schmidt:2010}.

As already mentioned that in the present analysis, the nonlinear evolution of dark matter overdensity and its clustering is studied with the assumption of spherical collapse. Spherical collapse is  a semi-analytic approach \cite{Gunn:1972sv,Liddle:1993fq,Padmanabhan:1999} to study the dynamics of matter overdensity. Evolution of spherical homogeneous overdensity is studied using the fully nonlinear equation derived from Newtonian hydrodynamics. The overdence region is assumed to be spherically symmetric with a uniform density of dark matter which is higher than the background density. It is considered as a closed sub-universe expanding with Hubble flow. But the expansion slows down and after reaching a maximum radius, it stars compression and eventually collapses due to gravitational attraction. Virialization of gravitational potential and thermal energy is introduced in the context of spherical collapse to explain the finite size of the collapsed object. The effective nature of dark energy can have its signature on the collapse of dark matter overdensities and consequently on the formation of large scale structure. The spherical collapse approach is numerically easier technique to apply in various dark energy models compared to the fully numerical simulations. There are several studies in this direction in the  literature \cite{Mota:2004pa,Nunes:2005fn,Mota:2008ne,He:2010ta,Pace:2010,Pace:2013pea,Basse:2010,Wint:2010,Devi:2010qp,Delliou:2012ik,Nazari-Pooya:2016bra,Setare:2017,Sapa:2018jja,Rajvanshi:2018xhf,Rajvanshi:2020das,Barros:2019hsk,Pace:2019vrs} where spherical collapse of matter overdensity is  investigated. Though the numerically sophisticated  approach to study the nonlinear evolution of cosmological perturbation and formation of large scale structure in the universe is the N-body simulation \cite{Jennings:2012,Maccio:2003yk,Baldi:2010vv,Boni:2011}. In case of geometric dark energy  the spherical collapse approach is also efficiently utilized to study the evolution dynamics of matter perturbation  \cite{Schafer:2008,Schmidt:2010}. Besides the evolution of matter density contrast, in the present work the number count of the collapsed objects in the DGP braneworld scenario is also emphasized. To obtain the number count of collapsed objects or the dark matter halos, two different mass function formalism has been adopted. Due to gravitational attraction the baryonic matter follows the distribution dark matter. Hence the  galaxy clusters are embedded in the dark matter halos. Thus the observed distribution of galaxy clusters provides the information about the distribution of dark matter halos in the universe. The cluster number count is an important observable of future observations of cosmic large scale structure. Study of number count of dark matter halos or galaxy clusters along redshift would be useful to test any model with upcoming cosmological observations.  

This paper is arranged as the following. In the next section (section \ref{DGPmodel}), the DGP braneworld cosmology in the FRW (Friedmann-Robertson-Walker) universe is discussed briefly. In section \ref{sphericalcollapse}, the linear and nonlinear evolution of matter overdensity and its gravitational collapse is discussed. In section \ref{clustercount} the number count of dark matter halos or the galaxy clusters are studied for the DGP cosmology and the results are compared with that of $\Lambda$CDM. Finally in section \ref{conclu}, it is concluded with a summarization of the results.

\section{DGP braneworld cosmology}
\label{DGPmodel}

\begin{figure}[tb]
\begin{center}
\includegraphics[angle=0, width=0.46\textwidth]{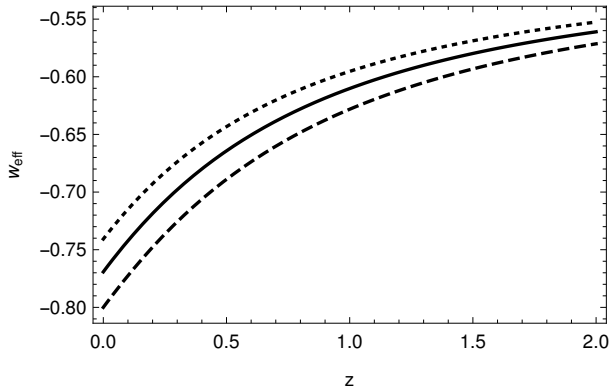}
\end{center}
\caption{{\small The effective equation of state parameter $w_{eff}(z)$ plot of the geometric dark energy in DGP braneworld cosmology. The parameters are fixed as $\Omega_{k0}=0.0$ and  $\Omega_{m0}=0.3$ (solid curve), $0.25$ (dashed curve), $0.35$ (dotted curve).}}
\label{fig:weff_a}
\end{figure}

The DGP braneworld model is based on the idea that our four dimensional universe lives on a five dimensional manifold \cite{Dvali:2000hr}. The gravitational interaction is modified due the presence of the five dimensional manifold. The modification in the gravitational interaction generates cosmic acceleration itself without introducing any exotic component in  the energy budget of the universe. In this framework, the first Friedmann equation is written as, 

\be
H^2+\frac{k}{a^2}-\frac{2M^3_{(5)}}{M^2_{(4)}}\left(H^2+\frac{k}{a^2}\right)^{\frac{1}{2}}=\frac{8\pi G}{3} \left(\rho_m + \rho_r \right).
\ee
As the gravitational constant is modified for 5D manifold, the Planck mass is also get modified. Here $M_{(5)}$ and $M_{(4)}$ are the 5D and 4D Planck masses. Finally, the scaled Hubble parameter for this model is obtained as, 

\be
h^2(a)=\frac{H^2(a)}{H_0^2}=\left[\sqrt{\Omega_{r0}a^{-4}+ \Omega_{m0}a^{-3}+\Omega_{rc}}+\sqrt{\Omega_{rc}}\right]^2 + \Omega_{k0}a^{-2},
\label{hubDGP}
\ee
where $r_c=M^2_{(4)}/2M^3_{(5)}$ called the crossover scale, $\Omega_{rc} = \frac{1}{4r_c^2H^2_0}$. Here the density parameters are defined as, $\Omega_{m0}=\frac{\rho_{m0}}{3H_0^2/8\pi G}$,  $\Omega_{r0}=\frac{\rho_{r0}}{3H_0^2/8\pi G}$, $\Omega_{k0}=-k/H_0^2$.
The normalized condition, $h(z=0)=1$ provides 
\be
\Omega_{rc}= \frac{(1- \Omega_{r0} - \Omega_{m0} - \Omega_{k0})^2}{4(1-\Omega_{k0})} .
\label{orc}
\ee
The dark energy density like contribution in the DGP braneworld model can be written as,

\be
\Delta H_{DE}^2=2\Omega_{rc}+2\sqrt{\Omega_{rc}}\sqrt{\Omega_{rc}+\Omega_{m0}a^{-3}+\Omega_{r0}a^{-4}}, 
\label{delH_DE}
\ee 
where $h^2(a)=\Omega_{m0}a^{-3}+\Omega_{r0}a^{-4}+\Omega_{k0}a^{-2}+\Delta H_{DE}^2$.  
At low redshift, neglecting the radiation contribution, the effective  equation of state parameter for the geometric dark energy model is given as,
\be
w_{eff}(a)=-1-\frac{1}{3}\frac{d\ln{\Delta H_{DE}^2}}{d\ln{a}},
\label{weff_eq}
\ee   
In figure \ref{fig:weff_a}, the plots of $w_{eff}(z)$ for the DGP braneworld cosmology are shown. The effective geometric dark energy equation of state evolves to  lower values at recent time generating the effective negative pressure required for the cosmic acceleration. 


\begin{figure}[tb]
\begin{center}
\includegraphics[angle=0, width=0.34\textwidth]{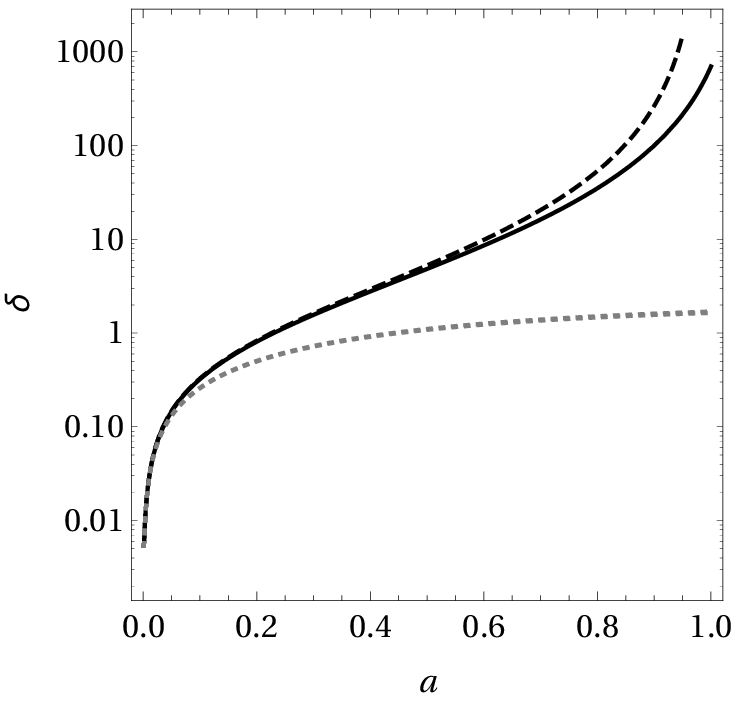}
\includegraphics[angle=0, width=0.34\textwidth]{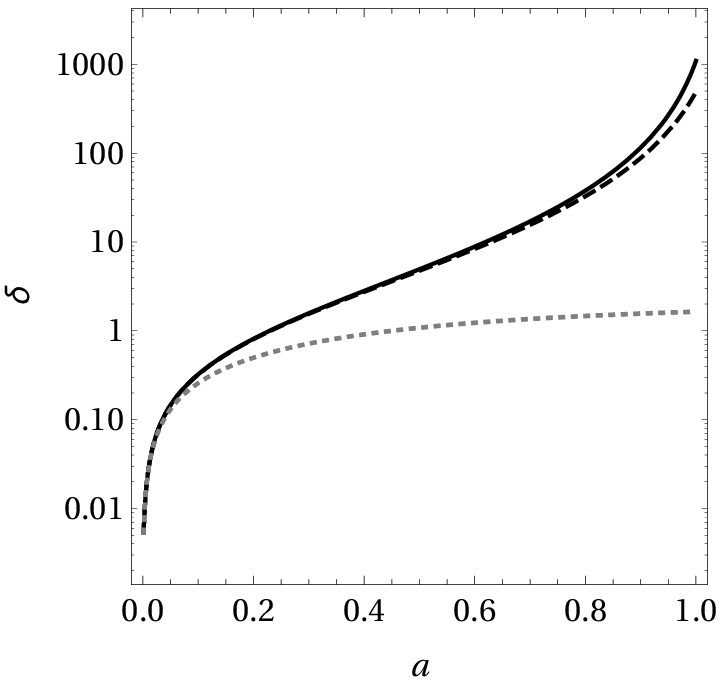}
\end{center}
\caption{{\small Linear (gray curves) and nonlinear (black curves) evolution of $\delta(a)$. In the left panel, the parameter values are $\Omega_{k0}=0$ and $\Omega_{m0}=0.3$ (black solid), $0.32$ (black dashed). The linear $\delta(a)$ in curve in left panel is degenerate for $\Omega_{m0}=0.3;~ 0.32$. In the right panel, the parameter values $\Omega_{m0}=0.3$ and $\Omega_{k0}=-0.01$ (black solid), $0.01$ (black dashed). The linear $\delta(a)$ curve in right panel is degenerate for $\Omega_{k0}=-0.01;~0.01$.}}
\label{fig:dela_plots}
\end{figure}

\section{Evolution of matter density contrast}
\label{sphericalcollapse}

Matter density contrast is defined as $\delta=\Delta\rho_m/\rho_m$ where $\Delta\rho_m$ is the deviation from homogeneous matter density $\rho_m$. The overdense region initially grows in size with Hubble expansion. Due to gravitational attraction, it gathers mass from the surrounding. After certain amount of mass accumulation, it starts collapsing. It is the fundamental process to form the large scale structure in the universe. The nonlinear evolution of the matter overdensities is important to understand the dynamics of gravitational collapse.  The semi-analytic approach, namely the spherical collapse model \cite{Gunn:1972sv,Liddle:1993fq,Padmanabhan:1999} is the simplest one to probe the nonlinear evolution of matter over density. It assumes that the overdense regions are spherically symmetric.  The nonlinear differential equation, that governs the time evolution of the matter density contrast is,  
\be
\ddot{\delta}+2H\dot{\delta}-4\pi G_{eff}\rho_m\delta(1+\delta)-\frac{4}{3}\frac{\dot{\delta}^2}{(1+\delta)}=0.
\label{del_nonlin}
\ee
The linear version of equation (\ref{del_nonlin}) is given as,
\be
\ddot{\delta}+2H\dot{\delta}-4\pi G_{eff}\rho_m\delta=0.
\label{del_lin}
\ee
The $G_{eff}$ is the effective gravitational constant for the braneworld model, given as, \cite{Chiba:2007rb},
{\small
\bea
\nonumber
\frac{G_{eff}(a)}{G}=\frac{4\Omega_m^2(a)-4(1-\Omega_k(a))^2+2\sqrt{1-\Omega_k(a)}(3}{3\Omega_m^2(a)-3(1-\Omega_k(a))^2+2\sqrt{1-\Omega_k(a)}(3}
\frac{-4\Omega_K(a)+2\Omega_m(a)\Omega_k(a)+\Omega^2_k(a))}{-4\Omega_K(a)+2\Omega_m(a)\Omega_k(a)+\Omega^2_k(a))},
\eea
}
or $G_{eff}=N(a)G$, where $G$ is the Newton's gravitational constant.
Using scale factor `a' as the argument of differentiation in equation (\ref{del_nonlin}) yield,  
\be
\delta''+\left(\frac{h'}{h}+\frac{3}{a}\right)\delta'-\frac{3\Omega_{m0}N(a)}{2a^5h^2}\delta(1+\delta)-\frac{4}{3}\frac{\delta'^2}{(1+\delta)}=0.
\label{del_nonlin_a}
\ee
Similarly the liner equation (eq. \ref{del_lin}) is  given as, 
\be
\delta''+\left(\frac{h'}{h}+\frac{3}{a}\right)\delta'-\frac{3\Omega_{m0}N(a)}{2a^5h^2}\delta=0.
\label{del_lin_a}
\ee
 
Equation (\ref{del_nonlin_a}) and (\ref{del_lin_a}) are studied numerically with the background given by equation (\ref{hubDGP}). In figure \ref{fig:dela_plots} the linear and nonlinear evolution of $\delta$ are shown with varying values of parameters ($\Omega_{m0},\Omega_{k0}$).  The initial conditions are fixed at $a_i=0.001$ and the initial values are fixed as $\delta_i=0.0054$ and $\delta'_i=0.0001$.  The linear evolution of $\delta$ is found to be very less sensitive to the change in parameter values, but the nonlinear evolution differs with variation in parameter values. The $\delta(a)$ curves are found to be indifferent with the change of the initial value $\delta'_i$.

\begin{figure}[tb]
\begin{center}
\includegraphics[angle=0, width=0.34\textwidth]{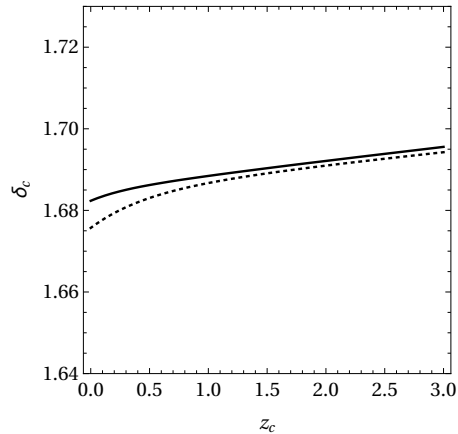}
\end{center}
\caption{{\small Critical density at collapse $\delta_c$ as a function of redshift of collapse $z_c$ for the DGP braneworld cosmology (solid curve),  and the LCDM cosmology (dotted curve).}}
\label{fig:delCz}
\end{figure}

Another important quantity to study the spherical collapse of dark matter overdensity is the critical density contrast ($\delta_c$). It is defined as the value of the linear density contrast at the redshift where the nonlinear density contrast diverges. Changing the initial condition $\delta_i$ in equation (equation (\ref{del_nonlin_a})) the redshift of nonlinear collapse can be changed. Thus the critical density contrast $\delta_c$ is obtained as a function of redshift at the collapse ($z_c$). The curves of $\delta_c(z_c)$ for the DGP braneworld cosmology is shown in figure \ref{fig:delCz}. For a comparison, the $\delta_c$ for $\Lambda$CDM cosmology is also shown. The critical density at collapse $\delta_c$ is important to study the distribution of galaxy cluster number or the number dark matter halos along redshift. The $\delta_c$ curves remain unchanged with small variation of the cosmological parameters. It is clearly due the less sensitivity of linear $\delta$ to the change of parameter values.

\section{Halo mass function and cluster number count}
\label{clustercount}

The gravitational collapse of the matter overdensity is the basic process of large scale structure formation in the universe. The objects, formed by the collapse, are called the dark matter halos. The baryonic  matter follows the distribution of dark matter. The idea is that the galaxy clusters are embedded in the dark matter halos.   Thus the observed distribution of galaxy clusters is the probe of dark matter halos in the universe. In this section, the number distribution of dark matter halos or the cluster number count is studied for the DGP braneworld cosmology using the spherical collapse model. In the semi-analytic approach, two different mathematical formulation of halo mass is used to evaluate the number count of collapsed objects or the galaxy clusters along the redshift. The first one is the Press-Schechter mass function \cite{Press:1973iz} and the other one, which is a generalization of the first one, is called the Sheth-Tormen mass function \cite{Sheth:1999mn}. The mathematical formulations of halo mass function are based on the assumption of a Gaussian distribution of the matter density field.  The comoving number density of collapsed object (galaxy clusters) at a certain redshift $z$ having mass range $M$ to $M+dM$ can be expressed as,
\be
\frac{dn(M,z)}{dM}=-\frac{\rho_{m0}}{M}\frac{d\ln{\sigma(M,z)}}{dM}f(\sigma(M,z)),
\ee
where $f(\sigma)$ is called the mass function. The mass function formula, proposed by Press and Schechter \cite{Press:1973iz}, is given as
\be
f_{PS}(\sigma)=\sqrt{\frac{2}{\pi}}\frac{\delta_c(z)}{\sigma(M,z)}\exp{\left[-\frac{\delta^2_c(z)}{2\sigma^2(M,z)} \right]}.
\label{fPS}
\ee
The $\sigma(M,z)$ is the corresponding rms density fluctuation in a sphere of radius $r$ enclosing a mass M. The $\sigma(M,z)$ can be expressed in terms of the linerised growth factor $g(z)=\delta(z)/\delta(0)$, and the rms of density fluctuation at a fixed length $r_8=8h_0^{-1}$Mpc as,

\be
\sigma(z,M)=\sigma(0,M_8)\left(\frac{M}{M_8}\right)^{-\gamma/3}g(z),
\ee
where $M_8=6\times10^{14}\Omega_{m0}h_0^{-1}M_{\odot}$, the mass within a sphere of radius $r_8$ and the $M_{\odot}$ is  the solar mass. The $\gamma$ is given as
\be
\gamma=(0.3\Omega_{m0}h_0+0.2)\left[2.92+\frac{1}{3}\log{\left(\frac{M}{M_8}\right)}\right].
\ee
Here $h_0$ is the Hubble constant $H_0$ scaled by $100km~s^{-1}Mpc^{-1}$.
The number of collapsed objects or cluster number count having mass range $M_i<M<M_s$ per redshift and square degree yield as,
\be
{\mathcal N}(z)=\int_{1deg^2}d\Omega\left( \frac{c}{H(z)}\left[\int_0^z\frac{c}{H(x)}dx\right]^2\right)\int_{M_i}^{M_s}\frac{dn}{dM}dM.
\label{numbercount_eq}
\ee

A general nature of cluster count is successfully depicted by the Press-Schechter fromalism, but it suffers from the prediction of higher abundance of galaxy cluster at low redshift and lower abundance of clusters at high redshift compared to the result obtained in simulation of dark matter halo formation \cite{Reed:2006rw}. To alleviate this issue, Sheth and Tormen \cite{Sheth:1999mn} proposed a modified version  of the mass function formula, which is given as,

\be
f_{ST}(\sigma)=A\sqrt{\frac{2}{\pi}}\left[1+\left( \frac{\sigma^2(M,z)}{a\delta^2_c(z)}\right)^p\right]\frac{\delta_c(z)}{\sigma(M,z)}\exp{\left[-\frac{a\delta_c^2(z)}{2\sigma^2(M,z)}\right]}.
\label{fST}
\ee
Three new parameters $(a,p,A)$ are introduced  by the Sheth-Tormen mass function formula, given in equation (\ref{fST}). For the values of the parameters $(a,p,A)$ as $(1,0,\frac{1}{2})$ the Sheth-Torman mass funtion mimics the Press-Schechter mass function. In the present work, while studying the distribution of cluster number count  using Sheth-Tormen mass function formula, the values of the parameter $(a,p,A)$ are fixed at $(0.707,0.3,0.322)$ as obtained form the simulation  of dark matter halo formation \cite{Reed:2006rw}. In the present work, the cluster number count is studied for the both the mass function formulas.

\begin{figure}[tb]
\begin{center}
\includegraphics[angle=0, width=0.3\textwidth]{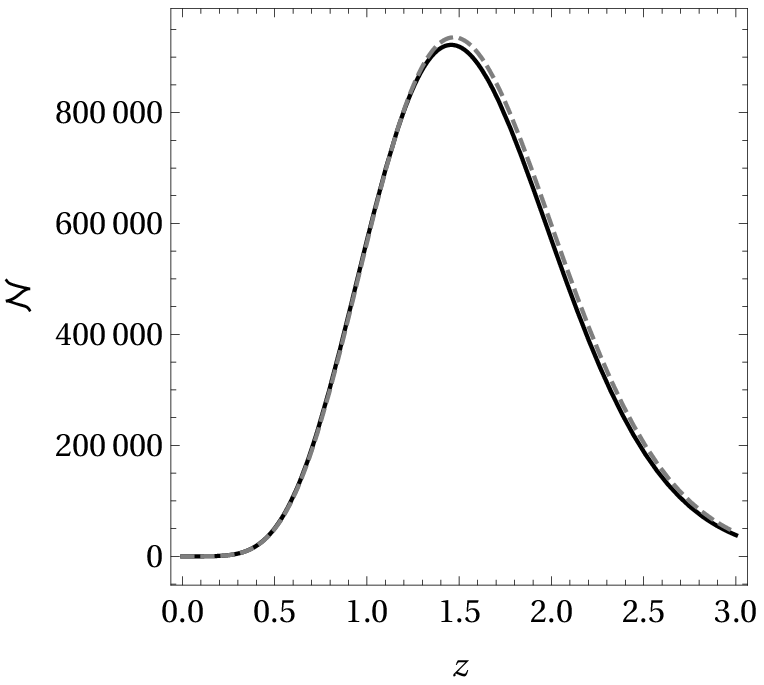}
\includegraphics[angle=0, width=0.3\textwidth]{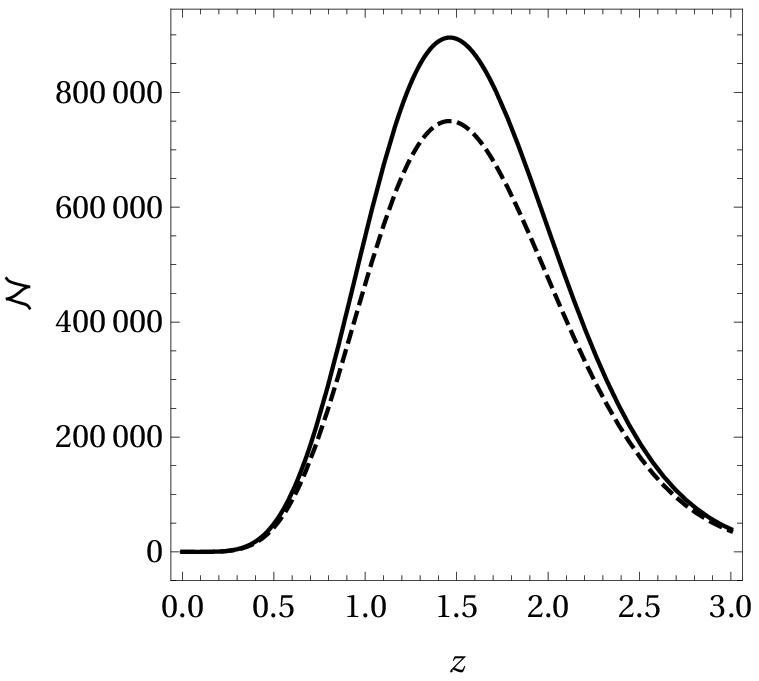}
\includegraphics[angle=0, width=0.3\textwidth]{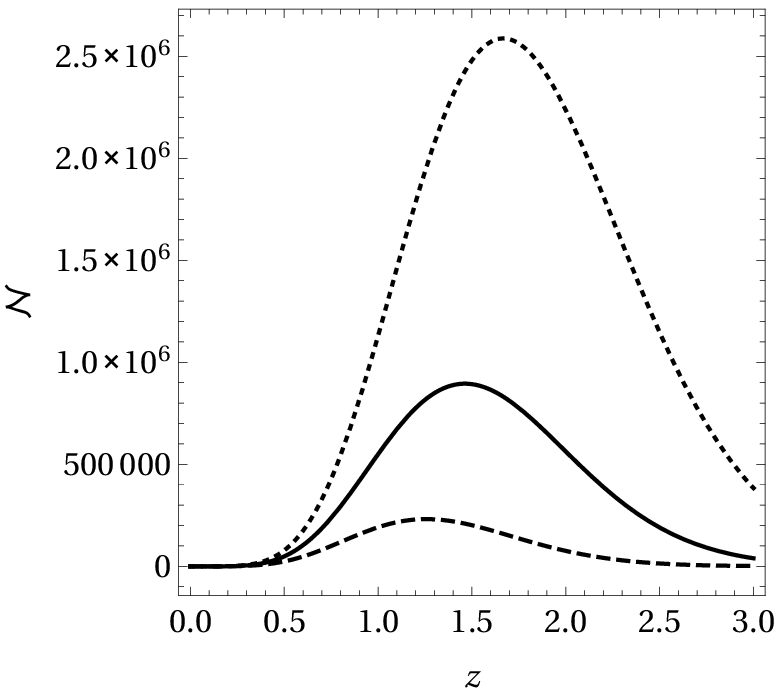}
\end{center}
\caption{{\small Cluster number count ${\mathcal N}(z)$ plots using Press-Schechter mass function for DGP braneworld cosmology. In the left panel, the parameters are fixed at $\Omega_{m0}=0.3$, $\sigma_8=0.8$ and $\Omega_{k0}=-0.01$ (solid curve), $\Omega_{k0}=0.01$ (dashed curve). In the middle panel, the parameters are fixed at $\Omega_{k0}=0.0$, $\sigma_8=0.8$ and $\Omega_{m0}=0.30$ (solid curve), $\Omega_{m0}=0.25$ (dashed curve). In the right panel, the parameters are fixed at $\Omega_{m0}=0.3$, $\Omega_{k0}=0.0$ and $\sigma_8=0.8$ (solid curve), $\sigma_8=0.7$ (dashed curve), $\sigma_8=0.9$ (dotted curve). }}
\label{fig:ClusterCount_PS}
\end{figure}
\begin{figure}[tb]
\begin{center}
\includegraphics[angle=0, width=0.3\textwidth]{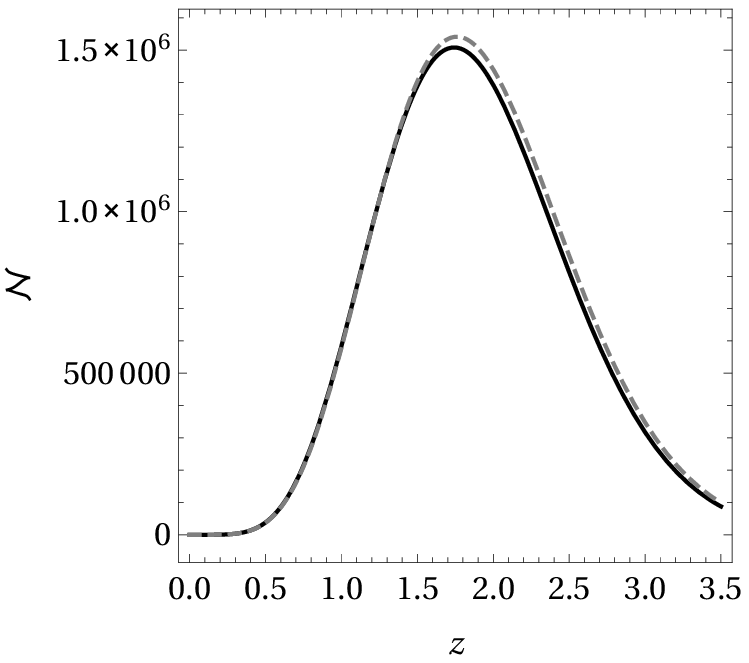}
\includegraphics[angle=0, width=0.3\textwidth]{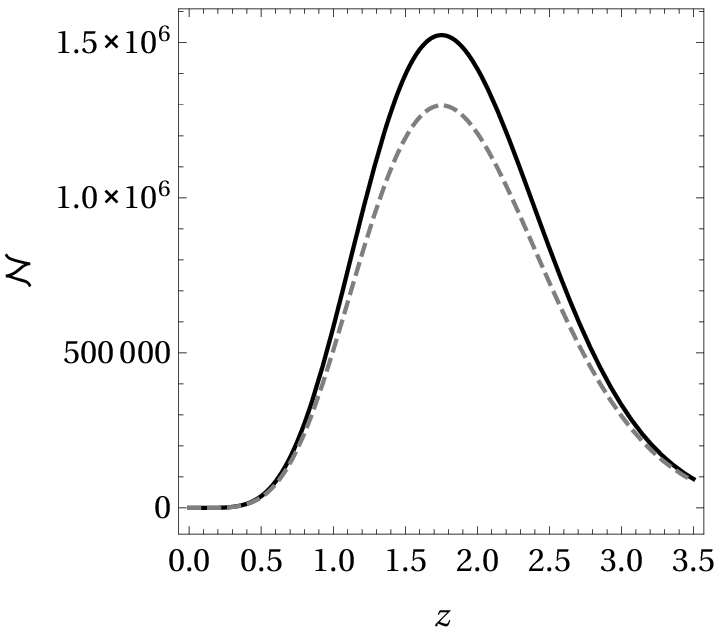}
\includegraphics[angle=0, width=0.3\textwidth]{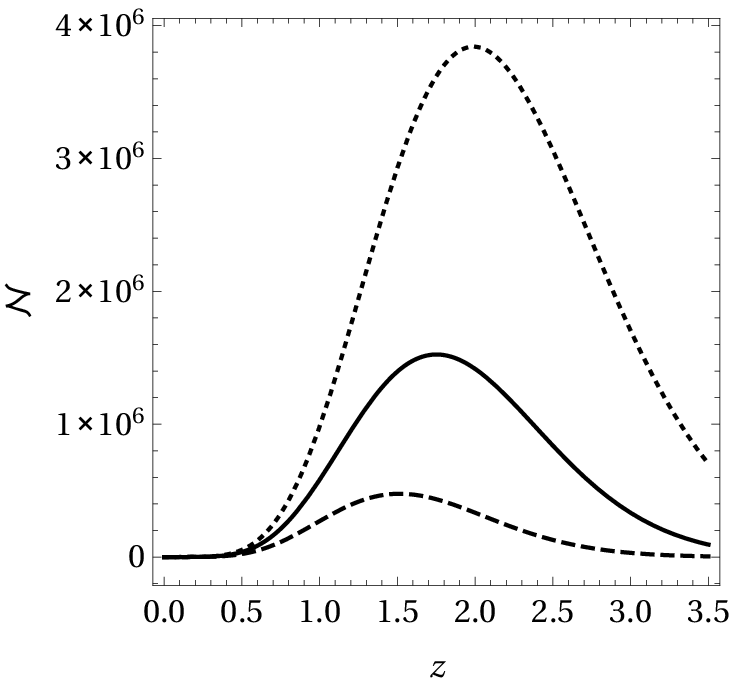}
\end{center}
\caption{{\small Cluster number count ${\mathcal N}(z)$ plots using Sheth-Tormen mass function for DGP braneworld cosmology. In the left panel, the parameters are fixed at $\Omega_{m0}=0.3$, $\sigma_8=0.8$ and $\Omega_{k0}=-0.01$ (solid curve), $\Omega_{k0}=0.01$ (dashed curve). In the middle panel, the parameters are fixed at $\Omega_{k0}=0.0$, $\sigma_8=0.8$ and $\Omega_{m0}=0.30$ (solid curve), $\Omega_{m0}=0.25$ (dashed curve). In the right panel, the parameters are fixed at $\Omega_{m0}=0.3$, $\Omega_{k0}=0.0$ and $\sigma_8=0.8$ (solid curve), $\sigma_8=0.7$ (dashed curve), $\sigma_8=0.9$ (dotted curve).}}
\label{fig:ClusterCount_ST}
\end{figure}

\begin{figure}[tb]
\begin{center}
\includegraphics[angle=0, width=0.34\textwidth]{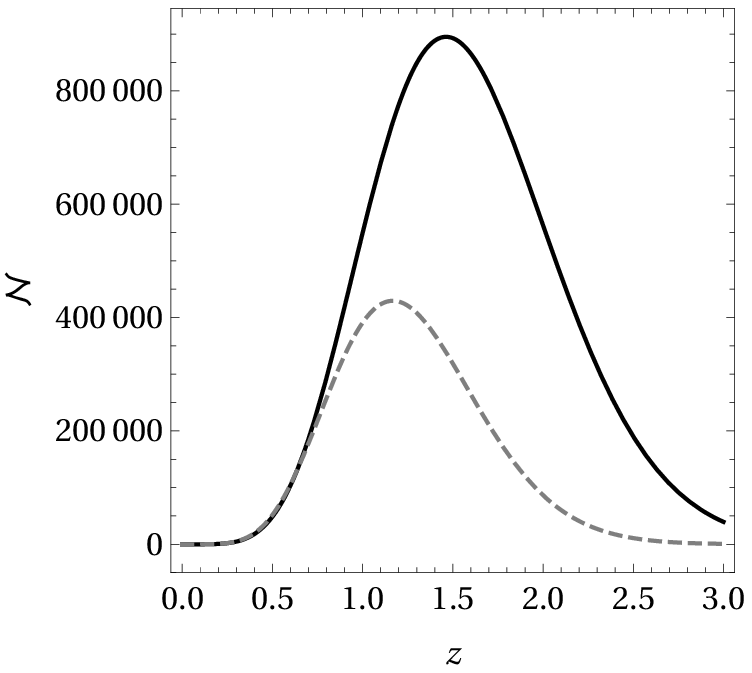}
\includegraphics[angle=0, width=0.34\textwidth]{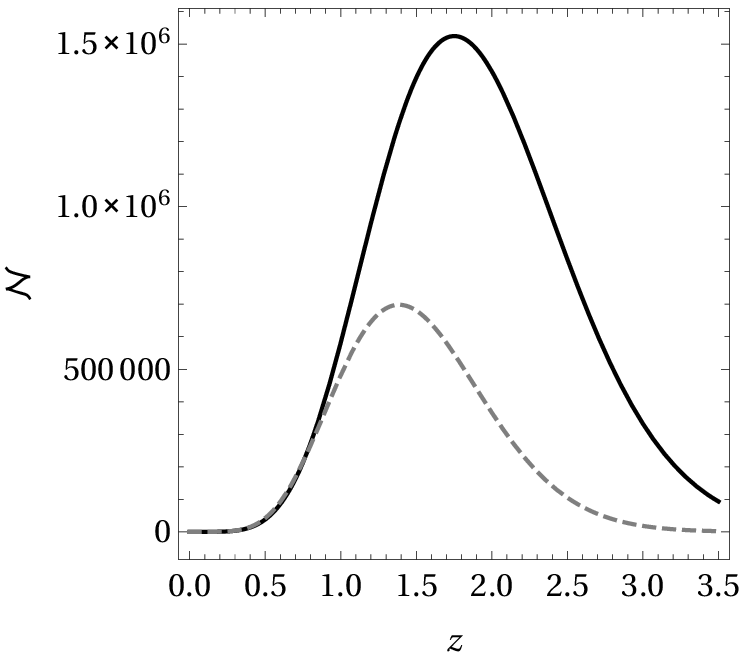}
\end{center}
\caption{{\small Cluster number count ${\mathcal N}(z)$ plots for DGP cosmology (solid curves) and for $\Lambda$CDM (dotted curve) with parameter values $\Omega_{m0}=0.3$, $\Omega_{k0}=0.0$ and $\sigma_8=0.8$. The left panel is obtained for the Press-Schechter mass function and right panel is obtained for Sheth-Tormen mass function.}}
\label{fig:CC_DGP_LCDM}
\end{figure}

The number count of dark matter halos or galaxy cluster for the Press-Schechter mass function are shown in figure \ref{fig:ClusterCount_PS}. The variation of cluster count with changing values of parameters $\Omega_{m0}$, $\Omega_{k0}$ and $\sigma_8$ are studied in figure \ref{fig:ClusterCount_PS}. In the left panel of figure \ref{fig:ClusterCount_PS}, ${\mathcal N}(z)$ plots are shown varying the value of $\Omega_{k0}$. The middle panel of figure \ref{fig:ClusterCount_PS} shows the ${\mathcal N}(z)$ plots varying the values of $\Omega_{m0}$. The right panel of figure \ref{fig:ClusterCount_PS} shows the ${\mathcal N}(z)$ plots for different values of the parameter $\sigma_8$. In the present study, the value of Hubble constant $H_0$ is fixed at $70~km~s^{-1}Mpc^{-1}$. Similarly figure \ref{fig:ClusterCount_ST} presents the cluster number count count plots for the Sheth-Tormen mass function with similar variation of the parameters. It is observed that the ${\mathcal N}(z)$ is less effected by the change of spatial curvature. On the other hand, it varies significantly with changing $\Omega_{m0}$. A higher value of $\Omega_{m0}$ produces higher number of dark matter halos. The cluster number count is also highly effected by  the rms fluctuation of matter density. Number count increases with the increase in the value of parameter $\sigma_8$. The effect of changing parameter values is very similar in case of Press-Schechter and  Sheth-Tormen mass function.

For a comparison with $\Lambda$CDM cosmology, the cluster number count ${\mathcal N}(z)$ plots for DGP braneworld model and $\Lambda$CDM model with same values of parameters are shown figure \ref{fig:CC_DGP_LCDM}. Parameters are fixed as $\Omega_{m0}=0.3$, $\Omega_{k0}=0.0$ and $\sigma_8=0.8$. The number count is found to be substantially enhanced in case of DGP.  As already mentioned, the lower value of $\sigma_8$ effectively decreases the number count, the value of $\sigma_8$ can be tuned for DGP model to produce number count of galaxy clusters similar to that of $\Lambda$CDM keeping other parameters same for both the models.

\section{Conclusion}
\label{conclu}
The present study deals with the cosmological implication of a higher dimensional gravity theory, namely the DGP braneworld model. The sherical collapse of matter overdensity has been investigated in the present context. The DGP braneworld is a 5-dimensional theory of gravity and at a length scale smaller than then the crossover length ($r_c$), the gravity is restricted on the 4-dimensional brane. The modification of gravity theory has its signature on the evolution of matter overdensity.  The signature of higher dimensional gravity theory on the evolution dynamics of matter density contrast comes through the modification of Friedmann equation governing the background as well as through the modification gravitational constant in the higher dimensional theory.    The linear and nonlinear evolutions (equation (\ref{del_nonlin}) and (\ref{del_lin})) of matter density contrast are studied numerically in the present work. It is totally based on the assumption of an isotropic collapse of the overdense region due to gravitational attraction. The curves of linear and nonlinear evolution of matter density contrast $\delta$ are shown in figure \ref{fig:dela_plots}. The effect of variation of cosmological parameters on the $\delta(a)$ curve are emphasized. The linear evolution is very less effected by the changes in the parameter values. The nonlinear evolution is non degenerate with the change of parameter values. Increase in the value of background matter density increases the clustering of dark matter. On the other hand, positive spatial curvature ($\Omega_k<0$) slightly increases the dark matter clustering compared to the negative spatial curvature ($\Omega_k>0$). Higher value of rms fluctuation of matter density produces higher number of dark matter halos. 

The critical density contrast in case of DGP remains slightly higher than the $\Lambda$CDM when the parameter values are kept same in both the cases. The initial conditions are fixed at the scale factor $a_i=0.001$, at the era after decoupling of CMB photon from baryonic matter. The number count of dark matter halos or galaxy clusters varies with the gravity theory. One prime endeavour of the present analysis was to check galaxy cluster number count in case of DGP cosmology. In figure \ref{fig:ClusterCount_PS}, the cluster count plots are shown for the Press-Schechter mass function formula. The effect of variation of the parameters are also investigated. It is observed that the positive spatial curvature has a slightly lower cluster number at high redshift compared to that in case of negative spatial curvature. Lower matter density decreases the cluster number. The variation of cluster number with the change in the rms fluctuation of matter density is also studied. The curves are highly effected by the change in the value of $\sigma_8$. In figure \ref{fig:ClusterCount_ST}, cluster number count for the Sheth-Tormen mass function is shown. The effects to variation of parameters in this case are same as in the case of Press-Schechter mass function. 

Further the cluster count distribution in the present scenario is compared to that of $\Lambda$CDM in figure \ref{fig:CC_DGP_LCDM}. The DGP brameworld model produces much higher cluster number at high redshift if the parameters are fixed at same values for DGP and $\Lambda$CDM. The redshift at which the cluster number is maximum is also different in this two cases. A slightly lower value of the rms of matter density fluctuation subsequently decreases the number of galaxy cluster and also drag the redshift of maximum cluster number to a lower value. Thus in DGP cosmology, the cluster number count consistent  with cosmological standard model can be achieved with a lower rms of matter fluctuation. Upcoming cosmological observations, like the South Pole Telescope (SPT), eROSITA,  would be a effective to check the viability of various dark energy models including the geometric dark energy scenarios through the precise observations of galaxy clusters number along the redshift.

\vskip 1.0 cm

\section*{Acknowledgment}
The author acknowledge the financial support from the Science and Engineering Research Board (SERB), Department of Science and Technology, Government of India through National Post-Doctoral Fellowship (NPDF, File no. PDF/2018/001859). The author would also like to thank Prof. Anjan A. Sen and Dr. Supriya Pan for useful discussions.


\vskip 1.50 cm
\begin{thebibliography}{}



\bibitem{Riess:1998cb} 
  A.~G.~Riess {\it et al.} [Supernova Search Team],
  Astron.\ J.\  {\bf 116}, 1009 (1998).
 


\bibitem{Perlmutter:1998np} 
  S.~Perlmutter {\it et al.} [Supernova Cosmology Project Collaboration],
  Astrophys.\ J.\  {\bf 517}, 565 (1999).



\bibitem{Schmidt:1998ys} 
  B.~P.~Schmidt {\it et al.} [Supernova Search Team],
  Astrophys.\ J.\  {\bf 507}, 46 (1998).



\bibitem{Suzuki:2012}
N. Suzuki et al., 
Astrophys. J. {\bf 746}, 85 (2012).


\bibitem{Crocce:2015xpb}
M.~Crocce \textit{et al.} [DES],
Mon. Not. Roy. Astron. Soc. \textbf{455}, 4301 (2016).

\bibitem{Agathe:2019vsu}
V.~de Sainte Agathe \textit{et al.},
Astron. Astrophys. \textbf{629}, A85 (2019).


\bibitem{Aghanim:2018eyx}
N.~Aghanim \textit{et al.} [Planck],
[arXiv:1807.06209 [astro-ph.CO]].



\bibitem{Dvali:2000hr} 
  G.~R.~Dvali, G.~Gabadadze and M.~Porrati,
  Phys.\ Lett.\ B {\bf 485}, 208 (2000).

\bibitem{Brax:2004xh}
P.~Brax, C.~van de Bruck and A.~C.~Davis,
Rept. Prog. Phys. \textbf{67}, 2183 (2004).


\bibitem{Bertolami:1999dp} 
  O.~Bertolami and P.~J.~Martins,
  Phys.\ Rev.\ D {\bf 61}, 064007 (2000).
  
\bibitem{Banerjee:2000mj} 
  N.~Banerjee and D.~Pav\'{o}n,
  Phys.\ Rev.\ D {\bf 63}, 043504 (2001).
  
  
\bibitem{Banerjee:2000gt} 
  N.~Banerjee and D.~Pav\'{o}n,
  Class.\ Quant.\ Grav.\  {\bf 18}, 593 (2001).
 
 
\bibitem{Sen:2000zk} 
  S.~Sen and A.~A.~Sen,
  Phys.\ Rev.\ D {\bf 63}, 124006 (2001).


\bibitem{Nojiri:2005vv}
S.~Nojiri, S.~D.~Odintsov and M.~Sasaki,
Phys. Rev. D \textbf{71}, 123509 (2005).


\bibitem{Comelli:2005tn}
D.~Comelli,
Phys. Rev. D \textbf{72}, 064018 (2005).

\bibitem{Andrew:2007xa}
K.~Andrew, B.~Bolen and C.~A.~Middleton,
Gen. Rel. Grav. \textbf{39}, 2061 (2007).






\bibitem{Capozziello:2003tk} 
  S.~Capozziello, S.~Carloni and A.~Troisi,
  Recent Res.\ Dev.\ Astron.\ Astrophys.\  {\bf 1}, 625 (2003).
 
\bibitem{Carroll:2003wy} 
  S.~M.~Carroll, V.~Duvvuri, M.~Trodden and M.~S.~Turner,
  Phys.\ Rev.\ D {\bf 70}, 043528 (2004).
  
\bibitem{Carroll:2004de} 
  S.~M.~Carroll, A.~De Felice, V.~Duvvuri, D.~A.~Easson, M.~Trodden and M.~S.~Turner,
  Phys.\ Rev.\ D {\bf 71}, 063513 (2005).
 
\bibitem{Nojiri:2003ft} 
  S.~Nojiri and S.~D.~Odintsov,
  Phys.\ Rev.\ D {\bf 68}, 123512 (2003).




\bibitem{Starobinsky:1980}
A. A. Starobinsky, Phys. Lett. B {\bf 91}, 99 (1980). 



\bibitem{Hu:2007nk}
W.~Hu and I.~Sawicki,
Phys. Rev. D \textbf{76}, 064004 (2007).




\bibitem{Pahwa:2011td} 
  I.~Pahwa, D.~Choudhury and T.~R.~Seshadri,
  JCAP {\bf 1109}, 015 (2011).
  
  
\bibitem{Alam:2016wpf} 
  U.~Alam, S.~Bag and V.~Sahni,
  Phys.\ Rev.\ D {\bf 95}, 023524 (2017).

\bibitem{Sadeghi:2014jua}
J.~Sadeghi, M.~Khurshudyan and H.~Farahani,
Int. J. Mod. Phys. D \textbf{25}, 1650108 (2016).

\bibitem{Ranjit:2011}
C. Ranjit, S. Chakraborty and  U. Debnath,
Int. J. Theor. Phys., {\bf 51}, 2180 (2012).

\bibitem{Liang:2010jy}
N.~Liang and Z.~H.~Zhu,
Res. Astron. Astrophys. \textbf{11}, 497 (2011).

\bibitem{Basilakos:2013ij}
S.~Basilakos and P.~Stavrinos,
Phys. Rev. D \textbf{87}, 043506 (2013).

\bibitem{Biswas:2018mxe}
M.~Biswas, S.~Ghosh and U.~Debnath,
Int. J. Geom. Meth. Mod. Phys. \textbf{16}, 1950178 (2019).


\bibitem{Ravanpak:2019zdg}
A.~Ravanpak and G.~F.~Fadakar,
Mod. Phys. Lett. A \textbf{34}, 1950105 (2019).



\bibitem{Koyama:2005kd}
K.~Koyama and R.~Maartens,
JCAP \textbf{01}, 016 (2006).



\bibitem{Multamaki:2003}
T.~Multamaki, E. Gaztanaga and M. Manera,
Mon. Not. R. Astron. Soc. {\bf 344}, 761 (2003).



\bibitem{Watts:2017lzm}
A.~L.~Watts, P.~J.~Elahi, G.~F.~Lewis and C.~Power,
Mon. Not. Roy. Astron. Soc. \textbf{468}, 59 (2017).


\bibitem{Song:2007wd}
Y.~S.~Song,
Phys. Rev. D \textbf{77}, 124031 (2008).


\bibitem{Cardoso:2007xc}
A.~Cardoso, K.~Koyama, S.~S.~Seahra and F.~P.~Silva,
Phys. Rev. D \textbf{77}, 083512 (2008).


\bibitem{Brax:2012}
P. Brax and P. Valageas,
Phys. Rev. D {\bf 86}, 063512 (2012).



\bibitem{Carroll:2006jn}
S.~M.~Carroll, I.~Sawicki, A.~Silvestri and M.~Trodden,
New J. Phys. \textbf{8}, 323 (2006).


\bibitem{Schafer:2008}
B. M. Schafer and K. Koyama,
Mon. Not. R. Astron. Soc. 385, 411–422 (2008).


\bibitem{Schmidt:2010}
F. Schmidt, W. Hu and M. Lima,
Phys. Rev. D {\bf 81}, 063005 (2010).




\bibitem{Gunn:1972sv} 
  J.~E.~Gunn and J.~R.~Gott, III,
  Astrophys.\ J.\  {\bf 176}, 1 (1972).


\bibitem{Liddle:1993fq} 
  A.~R.~Liddle and D.~H.~Lyth,
  Phys.\ Rept.\  {\bf 231}, 1 (1993).


\bibitem{Padmanabhan:1999}
 T. Padmanabhan, {\it Structure Formation in the Universe}
(Cambridge University Press, 1999).




\bibitem{Mota:2004pa} 
  D.~F.~Mota and C.~van de Bruck,
  Astron.\ Astrophys.\  {\bf 421}, 71 (2004).


\bibitem{Nunes:2005fn} 
  N.~J.~Nunes, A.~C.~da Silva and N.~Aghanim,
  Astron.\ Astrophys.\  {\bf 450}, 899 (2006).


\bibitem{Mota:2008ne} 
  D.~F.~Mota,
  JCAP {\bf 0809}, 006 (2008).

\bibitem{He:2010ta} 
  J.~H.~He, B.~Wang, E.~Abdalla and D.~Pavon,
  JCAP {\bf 1012}, 022 (2010).


\bibitem{Pace:2010}
 F. Pace, J.-C. Waizmann and M. Bartelmann,
 Mon.\ Not.\ Roy.\ Astron.\ Soc.\  {\bf 406}, 1865 (2010).

\bibitem{Pace:2013pea} 
  F.~Pace, L.~Moscardini, R.~Crittenden, M.~Bartelmann and V.~Pettorino,
  Mon.\ Not.\ Roy.\ Astron.\ Soc.\  {\bf 437}, 547 (2014).


\bibitem{Basse:2010}
T. Basse, O. E. Bjaelde and Y. Y. Y. Wong,
JCAP {\bf 1110}, 038 (2011).



\bibitem{Wint:2010}
N. Wintergerst and  V. Pettorino,
Phys. Rev. D {\bf 82}, 103516 (2010).


\bibitem{Devi:2010qp}
N.~Devi and A.~A.~Sen,
Mon. Not. Roy. Astron. Soc. \textbf{413}, 2371 (2011).

\bibitem{Delliou:2012ik} 
  M.~Le Delliou and T.~Barreiro,
  JCAP {\bf 1302}, 037 (2013).

\bibitem{Nazari-Pooya:2016bra} 
  N.~Nazari-Pooya, M.~Malekjani, F.~Pace and D.~M.~Z.~Jassur,
  Mon.\ Not.\ Roy.\ Astron.\ Soc.\  {\bf 458}, 3795 (2016).

\bibitem{Setare:2017}
  M. R. Setare, F. Felegary and F. Darabi,
  Phys. Lett. B {\bf 772}, 70 (2017).


\bibitem{Sapa:2018jja}
S.~Sapa, K.~Karwan and D.~F.~Mota,
Phys. Rev. D \textbf{98}, 023528 (2018).


\bibitem{Rajvanshi:2018xhf}
M.~P.~Rajvanshi and J.~Bagla,
JCAP \textbf{06}, 018 (2018).


\bibitem{Rajvanshi:2020das}
M.~P.~Rajvanshi and J.~Bagla,
[arXiv:2003.07647 [astro-ph.CO]].


\bibitem{Barros:2019hsk} 
  B.~J.~Barros, T.~Barreiro and N.~J.~Nunes,
  Phys.\ Rev.\ D {\bf 101}, 023502 (2020).


\bibitem{Pace:2019vrs}
F.~Pace and Z.~Sakr,
[arXiv:1912.12250 [astro-ph.CO]].



\bibitem{Jennings:2012}
E. Jennings, {\it Simulations of Dark Energy Cosmologies}
(Springer-Verlag Berlin Heidelberg, 2012).


\bibitem{Maccio:2003yk} 
  A.~V.~Maccio, C.~Quercellini, R.~Mainini, L.~Amendola and S.~A.~Bonometto,
  Phys.\ Rev.\ D {\bf 69}, 123516 (2004).

\bibitem{Baldi:2010vv} 
  M.~Baldi,
  Mon.\ Not.\ Roy.\ Astron.\ Soc.\  {\bf 411}, 1077 (2011).


\bibitem{Boni:2011}
     C. D. Boni, K. Dolag, S. Ettori, L. Moscardini, V. Pettorino and C. Baccigalupi,
     Mon.\ Not.\ Roy.\ Astron.\ Soc.\  {\bf 415}, 2758 (2011).







\bibitem{Chiba:2007rb} 
  T.~Chiba and R.~Takahashi,
  Phys.\ Rev.\ D {\bf 75}, 101301 (2007).







  
\bibitem{Press:1973iz} 
  W.~H.~Press and P.~Schechter,
  Astrophys.\ J.\  {\bf 187}, 425 (1974).
  
  
  
\bibitem{Sheth:1999mn} 
  R.~K.~Sheth and G.~Tormen,
  Mon.\ Not.\ Roy.\ Astron.\ Soc.\  {\bf 308}, 119 (1999).


\bibitem{Reed:2006rw} 
  D.~Reed, R.~Bower, C.~Frenk, A.~Jenkins and T.~Theuns,
  Mon.\ Not.\ Roy.\ Astron.\ Soc.\  {\bf 374}, 2 (2007).






  

\end{thebibliography}
\end{document}